\documentclass[epjCONF,draft]{svjour}
\usepackage[final]{graphicx}
\usepackage[varg]{txfonts} 
\usepackage[latin1]{inputenc}
\usepackage{color}
\usepackage{ulem}
\usepackage{booktabs}
\usepackage{units}
\usepackage[mathlines]{lineno}

\newcommand{\comment}[1]{ }

\session-title{Conference Title, to be filled}
\begin{document}


\title{Air Shower Simulation and Hadronic Interactions}

\author{
        Jeff Allen\inst{1},
        Antonella Castellina\inst{2},
        Ralph Engel\inst{3},
        Katsuaki Kasahara\inst{4},
        Stanislav Knurenko\inst{5},
        Tanguy Pierog\inst{3},
        Artem Sabourov\inst{5},
        Benjamin T. Stokes\inst{6},
        Ralf Ulrich\inst{3},\\[2mm]
        for the Pierre Auger, Telescope Array, and Yakutsk Collaborations, and\\[2mm]
        Sergey Ostapchenko\inst{7} and
        Takashi Sako\inst{8}
}

\institute{New York University, 4 Washington Place, New York, NY, USA
\and
Osservatorio Astrofisico di Torino (INAF), Universit\`a di Torino and Sezione INFN, Torino, Italy
\and
Karlsruhe Institute of Technology, Institut f\"{u}r Kernphysik, Karlsruhe, Germany
\and
RISE, Waseda University, Tokyo, Japan
\and
Yu.G. Shafer Institute of Cosmophysical Research and Aeronomy, Yakutsk, Russia
\and
High Energy Astrophysics Institute and Department of Physics and Astronomy, University of Utah, Salt Lake City, UT, USA
\and
NTNU, Institutt for fysikk, 7491 Trondheim, Norway
\and
Solar-Terrestrial Environment Laboratory, Nagoya University, Japan
}

\abstract{
The aim of this report of the Working Group on {\it Hadronic Interactions and Air Shower Simulation} is to give an overview of the status of the field, emphasizing open questions and a comparison of relevant results of the different experiments. It is shown that an approximate overall understanding of extensive air showers and the corresponding hadronic interactions has been reached. The simulations provide a qualitative description of the bulk of the air shower observables. Discrepancies are however found when the correlation between measurements of the longitudinal shower profile are compared to that of the lateral particle distributions at ground. The report concludes with a list of important problems that should be addressed to make progress in understanding hadronic interactions and, hence, improve the reliability of air shower simulations.
}

\maketitle


\section{Introduction}

After the first interaction of a cosmic ray particle of very high energy in the atmosphere a multitude of subsequent interactions, leading to particle multiplication, and decay processes give rise to a cascade of secondary particles called extensive air shower (EAS)~\cite{Auger:1939x1}. With the electromagnetic and weak interactions being well described by perturbative calculations within the Standard Model of Particle Physics, the limited understanding of strong interactions becomes the dominant source of uncertainties of shower predictions. Even though Quantum Chromodynamics (QCD) is the well-established and experimentally confirmed theory of strong interactions, only processes with very large momentum transfer can be predicted from first principles until now. It is still not possible to calculate the bulk properties of multiparticle production as needed for air shower simulation. Additional, simplifying assumptions as well as phenomenological and empirical parametrizations are needed to develop models for hadronic interactions describing various particle production processes. These additional assumptions need to be verified, parametrizations constrained, and parameters tuned by comparisons to accelerator data.

The simulation of extensive air showers forms one of the pillars on which the data analysis of modern experiments for ultra-high energy cosmic rays (UHECRs) rests. Improving the understanding and modeling of hadronic particle production is one of the important prerequisites for a reliable interpretation of UHECR data. Even though calorimetric techniques based on the measurement of fluorescence and Cherenkov light have been developed for a nearly model-independent energy determination of extensive air showers, determining the type of the primary particles and, in particular, estimating the mass of the primary particle can only be done with the help of sets of simulated reference showers.\footnote{An exception to this rule is the detection of direct Cherenkov light emitted before a particle interacts in the atmosphere~\cite{Kieda:2000ky}. This technique is, however, not applicable at the highest energies.}

Much effort has been devoted to improving both the simulation of extensive air showers in the atmosphere and the understanding of the corresponding, and typically very complex detector response function.
In the following we will concentrate on discussing the simulation of extensive air showers as this aspect of the overall data simulation chain is the same in all UHECR experiments. We will assume that uncertainties arising from simulating the detector response due to the shower particles are much smaller and can be disregarded here. This is certainly not guaranteed per se and it is the task of each collaboration to verify the quality of the detector simulation before attributing possible discrepancies to, for example, shortcomings in the understanding of air showers or hadronic multiparticle production.

After recalling some basic features of air showers (Sec.~\ref{sec:ShowerPhysics}) we will give an overview of the most frequently used code packages for air shower simulation in Sec.~\ref{sec:ShowerCodes}. The LHC data allow us to test the hadronic interaction models employed in these code packages, for the first time, at equivalent energies beyond that of the knee in the cosmic ray spectrum. In Sec.~\ref{sec:LHC} some representative model predictions are compared to LHC data and implications are discussed. A good overall bracketing of the LHC data by model predictions is found, even though each of the models will have to be improved to obtain a satisfactory description of the LHC data. Therefore it is not surprising that the distributions of most of the shower observables are well reproduced by shower simulations (Sec.~\ref{sec:ShowerDescription}).

Taking advantage of the hybrid detection setups of the latest generation of UHECR detectors, longitudinal and lateral particle distributions can be measured shower-by-shower. A comparison of the particle densities at ground with the ones expected according to the shower longitudinal profile reveals possible shortcomings of the shower predictions that are not yet understood. The overall data set of such comparisons is discussed in Sec.~\ref{sec:Discrepancies} and the results from the Auger, TA, and Yakutsk Collaborations are compared. There are strong indications that a significant part of the shortcomings of the shower predictions are related to the muonic shower component. Therefore we summarize recent developments aiming at better understanding muon production in air showers in Sec.~\ref{sec:MuonProduction}.

Given the increased awareness of apparent deficits of current air shower predictions and related theory developments, the wealth of new data from the LHC, and the high-quality measurements of the current generation of air shower observatories, significant progress can be expected in the reliability of air shower predictions in the next few years. Open questions that need to be addressed to optimally benefit from these developments to improve the interpretation of EAS data and to solve some outstanding problems will be listed in the concluding section of this report.


\section{Physics of Air Showers\label{sec:ShowerPhysics}}

Reviews of the physics of extensive air showers and hadronic interactions can be found in, for example, \cite{Gaisser:1978kx,Knapp:2002vs,Anchordoqui:2004xb,Engel:2011zz} and a very instructive extension of Heitler's cascade model to hadronic showers is given in~\cite{Matthews:2005sd}. Here we give only a qualitative introduction to some aspects of shower physics that will be needed in the discussions later on.

\begin{figure}[htb!]
\begin{center}
\includegraphics[width=0.6\textwidth]{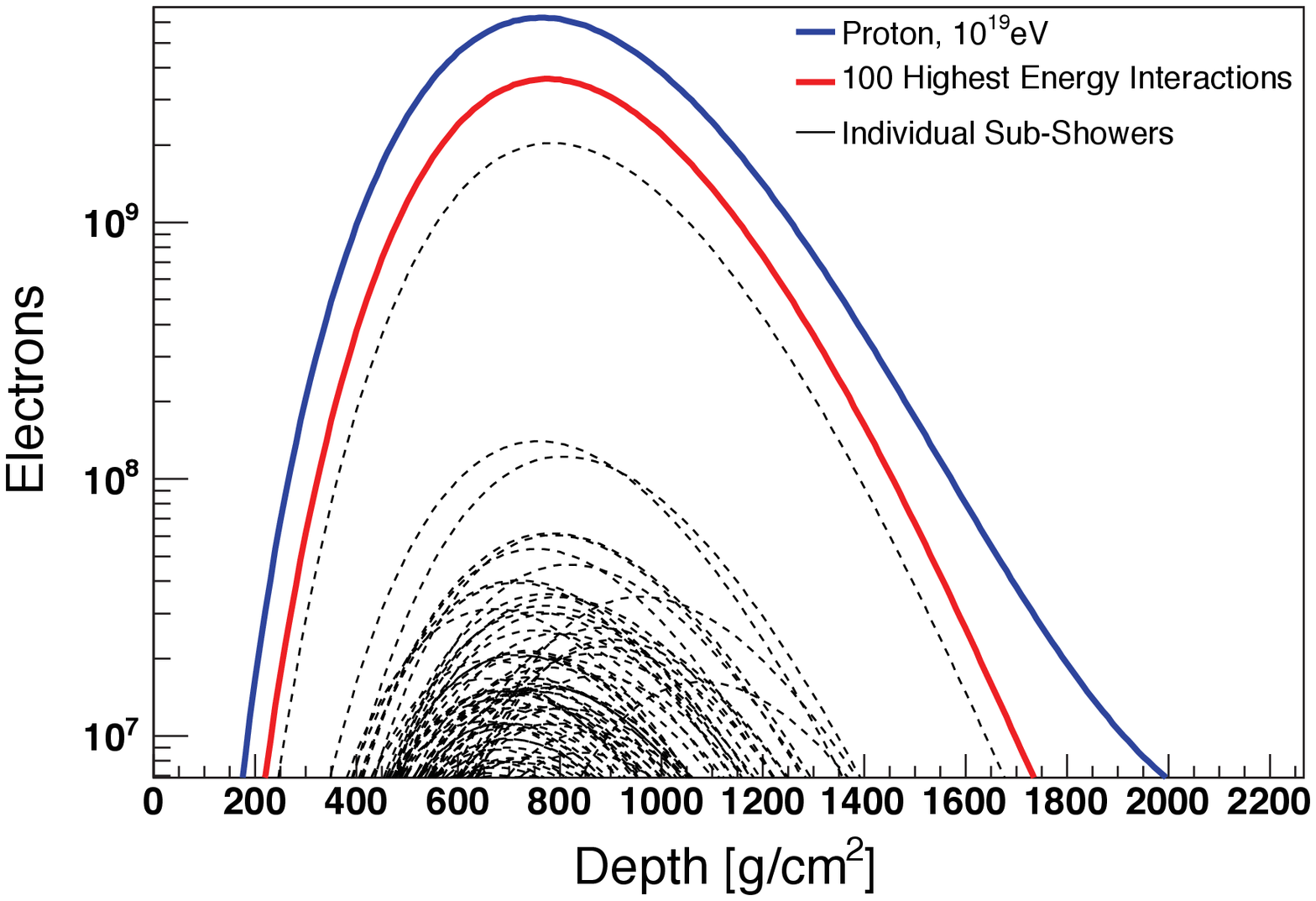}\\
\includegraphics[width=0.6\textwidth]{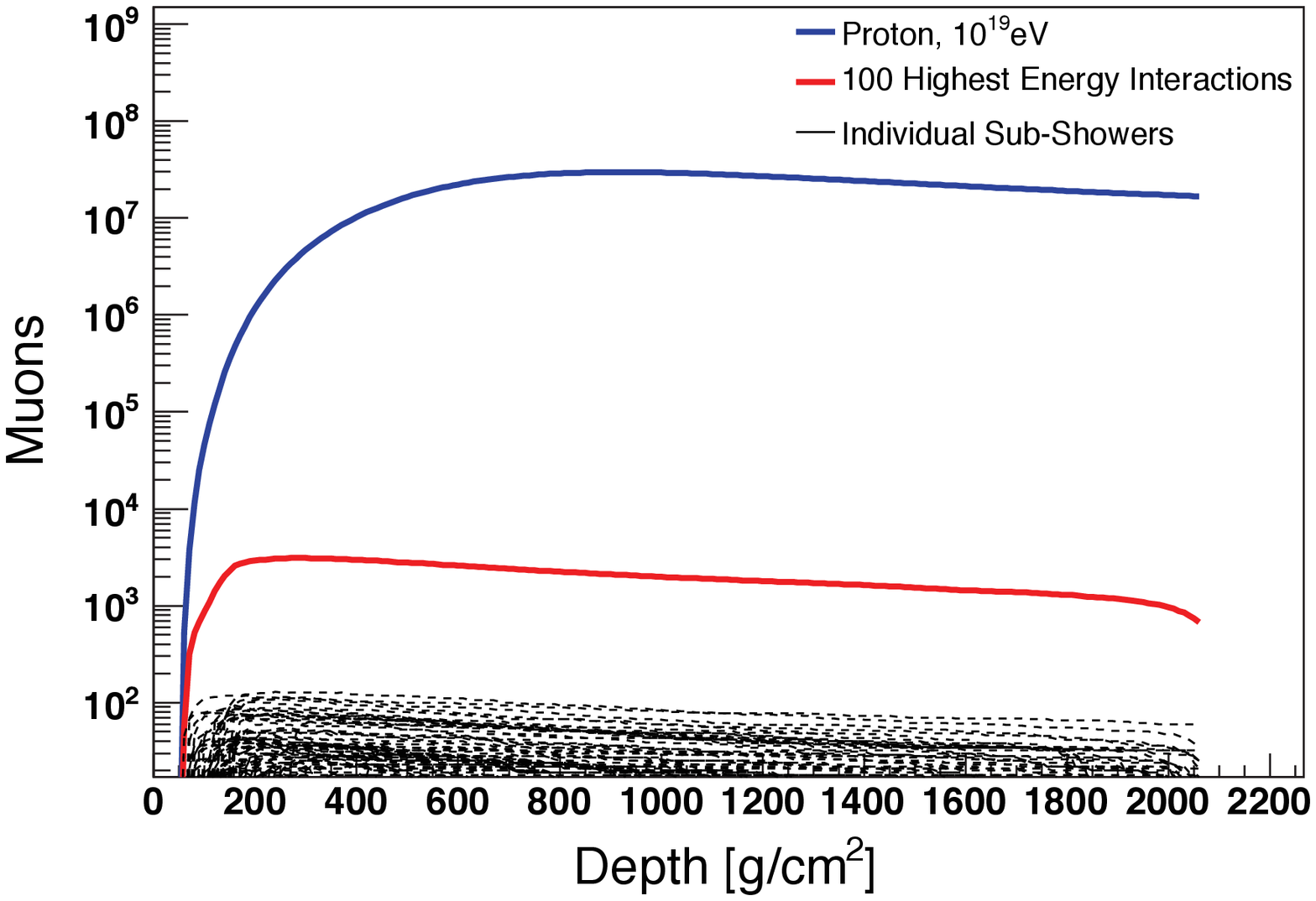}
\end{center}
\caption{Longitudinal shower profiles of electrons and muons. Shown are also the sub-showers produced by the secondary particles of the first 100 highest-energy interactions and the cumulative profile of these sub-showers~\cite{Ulrich-MPI:2011x1}. The simulation was done with a modified version of CONEX~\cite{Bergmann:2006yz} for a proton shower of $10^{19}$\,eV.
}
\label{fig:First100}
\end{figure}

With pions being the most abundant secondary particles of hadronic interactions at high energy, the difference in lifetime and decay products of neutral and charged pions lead to fundamentally different ranges of interaction energies that give rise to the electromagnetic and muonic shower components. In air showers, neutral pions decay almost always before interacting again (except $E_{\pi^0} > 10^{19}$\,eV) and provide high-energy photons that feed the electromagnetic shower component. In contrast, the energy of charged pions has to be degraded down to some $30 - 100$\,GeV before the probability to decay will exceed that of another interaction. In first approximation, the depth profile of charged shower particles, being dominated by $e^+ e^-$ of the electromagnetic component, is linked to the secondary particles of the first few interactions in the hadronic core of the shower. In contrast, the muons at ground stem from a chain of $8-10$ successive hadronic interactions  with the energy of the last interaction producing pions or kaons that lead to observed muons being in the range of $20-200$\,GeV~\cite{Drescher:2002vp,Meurer:2005dt}. This is illustrated in Fig.~\ref{fig:First100} showing the sub-showers produced by the secondary particles of the first $100$ highest-energy interactions in an air shower~\cite{Ulrich-MPI:2011x1}. While more than $50$\% of the electromagnetic shower particles are produced by decaying neutral pions of the first $100$ interactions in an air shower, only a negligible fraction of the muonic component stems from decaying charged pions and kaons of the same interactions.

The longitudinal profile of showers as well as the energy and angular distributions of charged particles in a shower exhibit universality features~\cite{Hillas:1982vn,Giller:2004cf,Giller:2005gp,Nerling:2006yt,Lafebre:2009en,Lipari:2008td} if considered as a function of shower age, here approximated by
\begin{equation}
s = \frac{3 X}{X + 2 X_{\rm max}}\ ,
\end{equation}
with $X$ being the slant depth along the shower axis and $X_{\rm max}$ the depth of shower maximum. Universality features are routinely used to estimate the Cherenkov light signal of showers, see, for example,~\cite{Nerling:2006yt,Baltrusaitis:1985mx}. Not only the electromagnetic particles but also the production depth as well as the energy and transverse momentum distributions of muons can be described by functions~\cite{Andringa:2011ik,Cazon:2012ti} that depend only weakly on the assumed hadronic interaction model. Breaking down the overall shower signal at ground into different muonic and electromagnetic components, the description of showers can be improved considerably \cite{Schmidt:2007vq,Schmidt:2009ge,Ave:2011x1}.

While the bulk of charged particles is considered for longitudinal profiles of the electromagnetic and muonic shower components, lateral particle distributions at ground are only measured at large distances from the shower axis. Particle densities at, for example, $600$ or $1000$\,m from the shower core are only indirectly related to the overall particle multiplicity of a shower at ground. This is reflected in the fact that, for example, the last interaction for producing a muon observable at ground has a median energy of $\sim 100$\,GeV for the KASCADE array (typical primary energy $3\times 10^{15}$\,eV; lateral distance $40-200$\,m, see~\cite{Meurer:2005dt}) and only $20 - 30$\,GeV for the Auger array (typical primary energy $10^{19}$\,eV; lateral distance $1000$\,m, see~\cite{Maris:2009x1}). Still universality arguments can also be applied to particle densities far away from the core~\cite{Schmidt:2007vq} but the model-related differences between the predictions are larger. In particular, low-energy interactions can lead to a model-dependent violation of universality profiles that is reflected in both the muonic and electromagnetic shower content at large lateral distance~\cite{Lafebre:2009en,Ave:2011x1}.


\section{Shower Simulation Packages and Hadronic Interaction Models\label{sec:ShowerCodes}}

The most frequently used code packages for the simulation of air showers for the Auger, TA and Yakutsk experiments are AIRES~\cite{Sciutto:AIRES}, CORSIKA~\cite{Heck98a}, and COSMOS~\cite{cosmos}. In these packages, the Monte Carlo method is applied to simulate the evolution of air showers with all secondary particles above a user-specified energy threshold. The large number of secondary particles that would have to be followed in the simulation of ultra-high energy showers makes the simulations so time consuming that thinning techniques are applied. Only a representative set of particles is included in the simulation below a pre-defined energy threshold. The discarded particles are accounted for by increasing the statistical weight of the particles remaining in the simulation~\cite{Hillas81a,Kobal:2001jx}. The main drawbacks of this method are artificial fluctuations related to particles with large statistical weight~\cite{Gorbunov:2007vj,Hansen:2010uk} and the need for treating weighted particles in the detector simulations. The Auger and TA Collaborations have developed different methods for de-thinning simulated showers, see \cite{Billoir:2008zz} and \cite{Stokes:2011wf}, respectively. Also a limited number of fully simulated showers of the highest energies are available in each collaboration (see, for example, \cite{Stokes:2011xk}) but the statistics is very limited and there is a lack of methods to work with such large amounts of data in an efficient way.

The classic Monte Carlo codes are complemented by hybrid simulation programs that combine the Monte Carlo technique with either numerical solutions of cascade equations (CONEX~\cite{Bergmann:2006yz} and
SENECA~\cite{Drescher:2002cr}) or libraries of pre-simulated air showers (COSMOS~\cite{cosmos}). The advantage of the hybrid codes is the much reduced CPU time needed for simulating showers at very high energy and the possibility to simulate showers either without thinning or with very low statistical weights. But there is always a remaining risk that rare shower fluctuations or local correlations might not be correctly described.

Different codes are applied for the simulation of the electromagnetic shower component. The code used in AIRES is based on the one developed by Hillas for MOCCA~\cite{Hillas81a,Hillas:1997tf}. CONEX, CORSIKA, and SENECA are interfaced to modified versions of EGS4~\cite{egs4}. Similar to AIRES, COSMOS comes with a custom-developed code for electromagnetic interactions called EPICS~\cite{cosmos}. The suppression of high-energy interactions of photons and electrons due to the Landau-Pomeranchuk-Migdal (LPM) effect (for example, see \cite{Cillis:1998hf}) is taken into account in all code packages.

Due to the different underlying approaches for modeling hadronic interactions at low ($E < 80 - 200$\,GeV) and high energies typically always two hadronic interaction models are employed in air shower simulations. The low-energy models GHEISHA~\cite{Fesefeldt85a}, FLUKA~\cite{Ferrari:2005zk}, and
UrQMD~\cite{Bleicher99a} are available in CORSIKA while AIRES and COSMOS employ custom-made codes.
Typical high-energy interaction models are DPMJET II~\cite{Ranft:1994fd} and
III~\cite{Roesler02a,Bopp:2005cr},
EPOS~\cite{Werner:2005jf,Pierog:2009zt,Werner:2010ss}, QGSJET
01~\cite{Kalmykov:1997te,Kalmykov:1993qe} and
II~\cite{Ostapchenko:2005nj,Ostapchenko:2004ss,Ostapchenko:2007qb},
and SIBYLL~\cite{Ahn:2009wx,Engel:1992vf,Fletcher:1994bd}.
A compilation of simulation results
obtained with many of these low- and high-energy models can be found
in~\cite{Heck:2002yf,Heck:2004rq,Heck08CORSIKASchool}.

\begin{figure}[htb!]
\begin{center}
\includegraphics[width=0.6\textwidth]{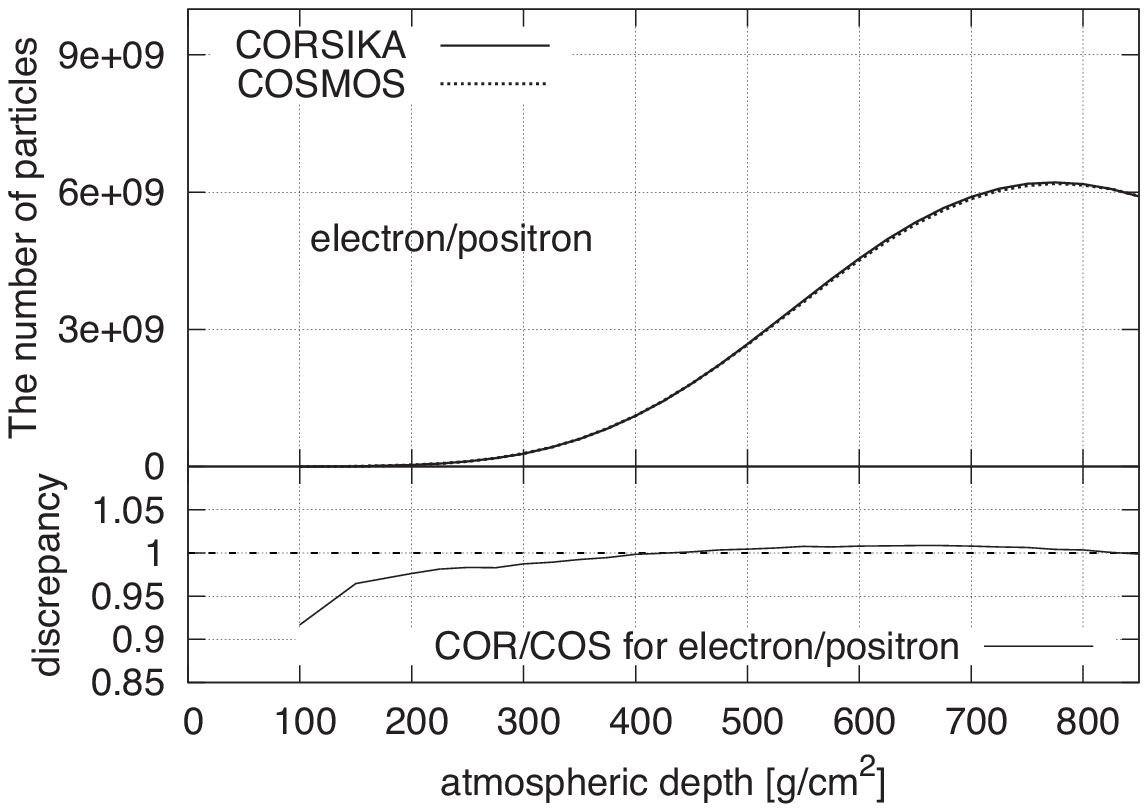}\\
\includegraphics[width=0.6\textwidth]{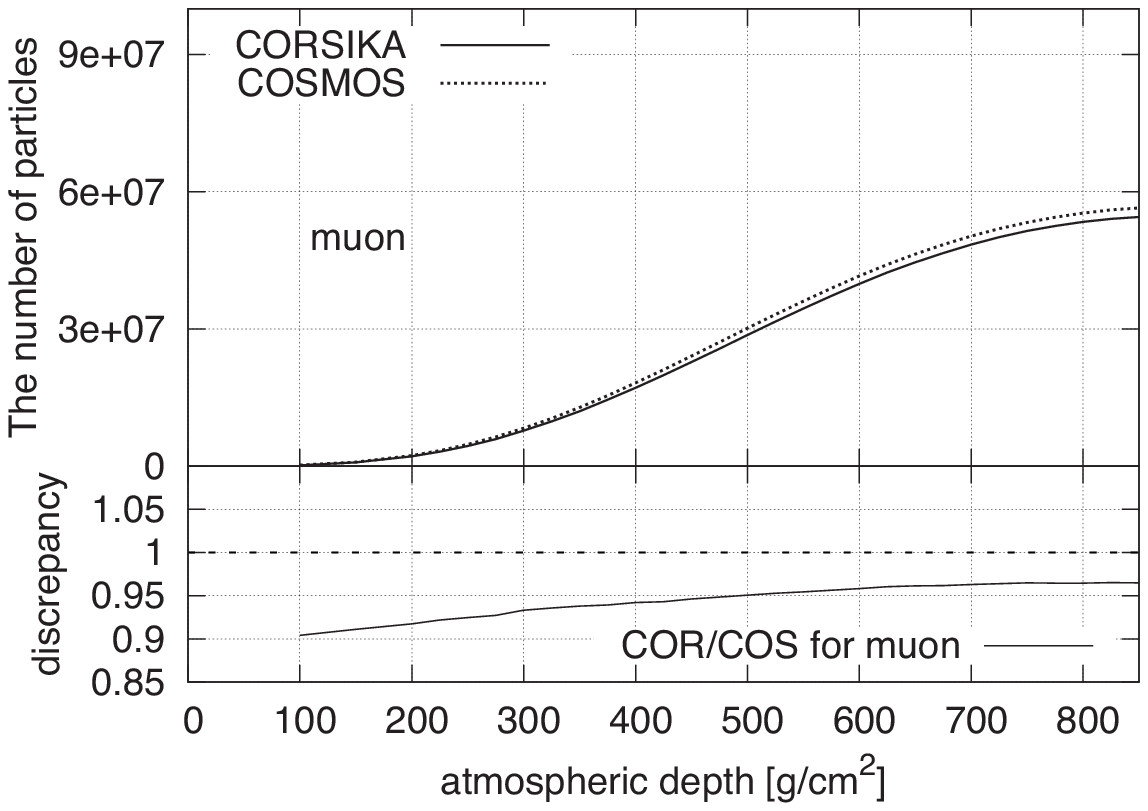}
\end{center}
\caption{Longitudinal shower profiles of electrons and muons. Predictions calculated with CORSIKA and COSMOS are compared for vertical proton showers of $10^{19}$\,eV~\cite{Kim:2011x1,Roh:2013ev}. High-energy interactions were simulated with QGSJET II.03 for energies above $80$\,GeV.
}
\label{fig:CorCosComp}
\end{figure}
Several comparisons of the predictions of shower simulation packages for the same interaction models are available in the literature, see \cite{Knapp:2002vs} and \cite{Ortiz:2004gb,Allen:2009x1,Kim:2011x1,Roh:2013ev,ToderoPeixoto:2013x1}.
In the ideal case, the predictions for air showers of a given energy and primary particle depend only on the user-selected energy thresholds and on the choice of hadronic interaction models selected for the simulation. While it is possible to chose a high-energy interaction model that is supported in all shower simulation codes, less flexibility is available for the low-energy interaction models, which are often different in each of the simulation packages. Therefore the results of these comparisons typically show a good agreement of the longitudinal shower profiles, which are mainly related to hadronic interactions at high energy. Differences are found for observables that are sensitive to low-energy interactions, which are, for example, the total number of muons and particle densities at large lateral distances. The differences between the predictions of different simulation packages are typically of the order of 5\%, increasing in some phase space regions up to $\sim 10$\%. For example, longitudinal shower profiles from a recent comparison between CORSIKA and COSMOS \cite{Kim:2011x1,Roh:2013ev} are shown in Fig.\ref{fig:CorCosComp}.


\section{Performance of Hadronic Interaction Models\label{sec:LHC}}

\begin{figure}[htb!]
\begin{center}
\includegraphics[width=0.7\textwidth]{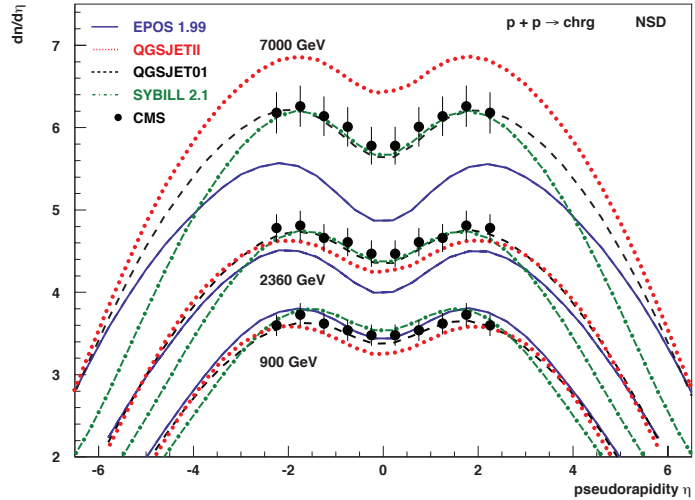}
\end{center}
\caption{Pseudorapidity distribution of charged particles measured at LHC. CMS data~\cite{Khachatryan:2010us} are shown together with model predictions. Similar data sets are available from the ATLAS and ALICE Collaborations~\cite{Aad:2010ac,Aamodt:2010pp,Aamodt:2010ft} (not shown). Below 2 TeV center-of-mass energy the models were tuned to previously available collider data from Tevatron and SPS.
}
\label{fig:EtaLHC}
\end{figure}

Most of the interaction models used in air shower simulations are not commonly applied in high energy physics (HEP) simulations and, conversely, HEP models are not used for air shower simulations. This is related to the fact that HEP models are typically limited to a set of primary particles that are available in accelerator experiments and optimized only for collider energies. Similarly, cosmic ray interaction models are not routinely used by HEP collaborations for acceptance correction calculations including the comparison of raw, uncorrected data as this can be done with the HEP event generators as well. Furthermore dedicated HEP event generators describe high-$p_\perp$ and electroweak physics processes in much more detail and contain many more parameters for improving the description of particular distributions than any of the cosmic ray or general purpose models. In contrast, interaction models for cosmic ray physics are tuned to describe data of accelerator measurements over a wide range of energies, often sacrificing a perfect description of some distributions in favour of a good overall reproduction of the whole data set.

\begin{figure}[htb!]
\begin{center}
\includegraphics[width=0.4\textwidth]{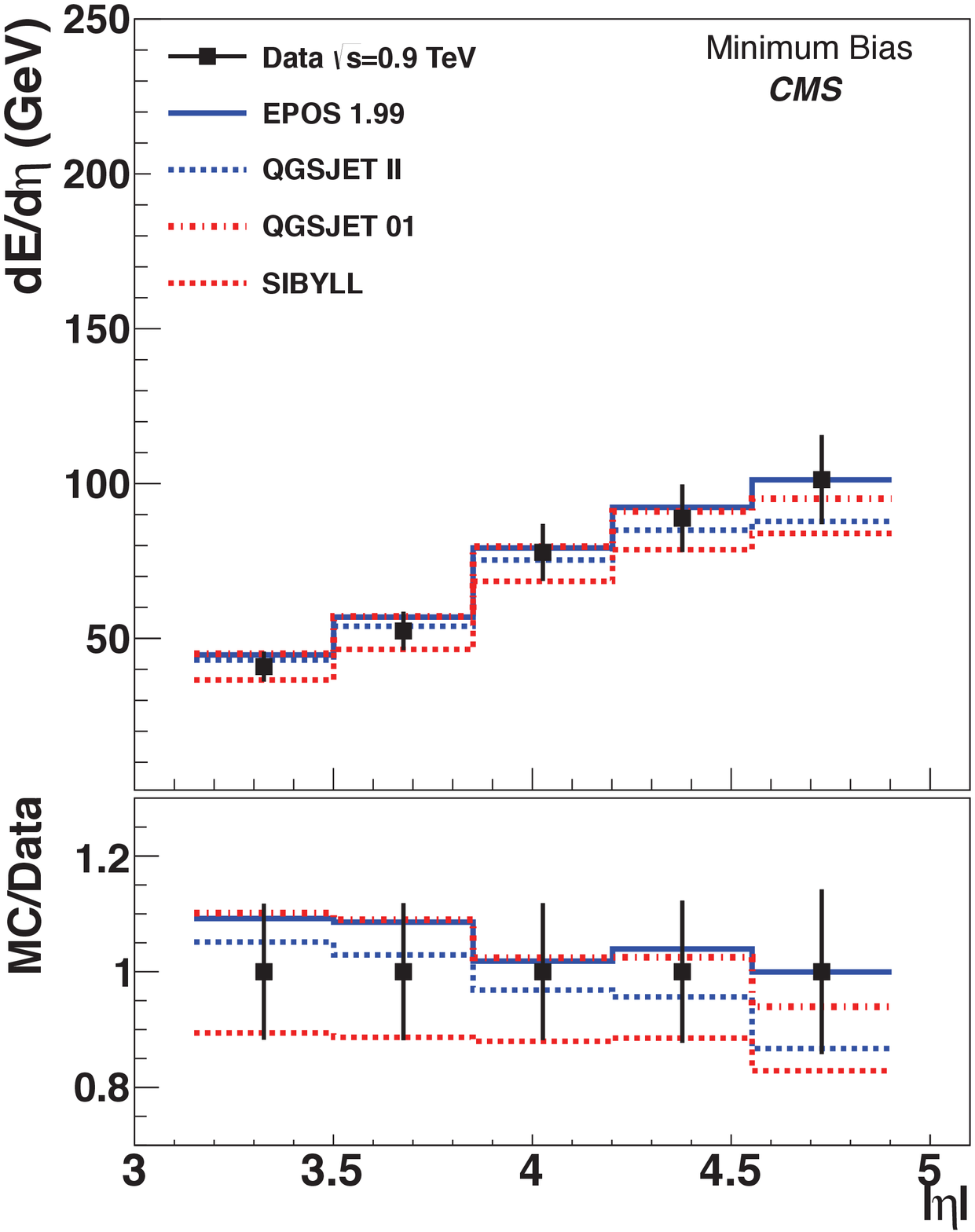}
\includegraphics[width=0.4\textwidth]{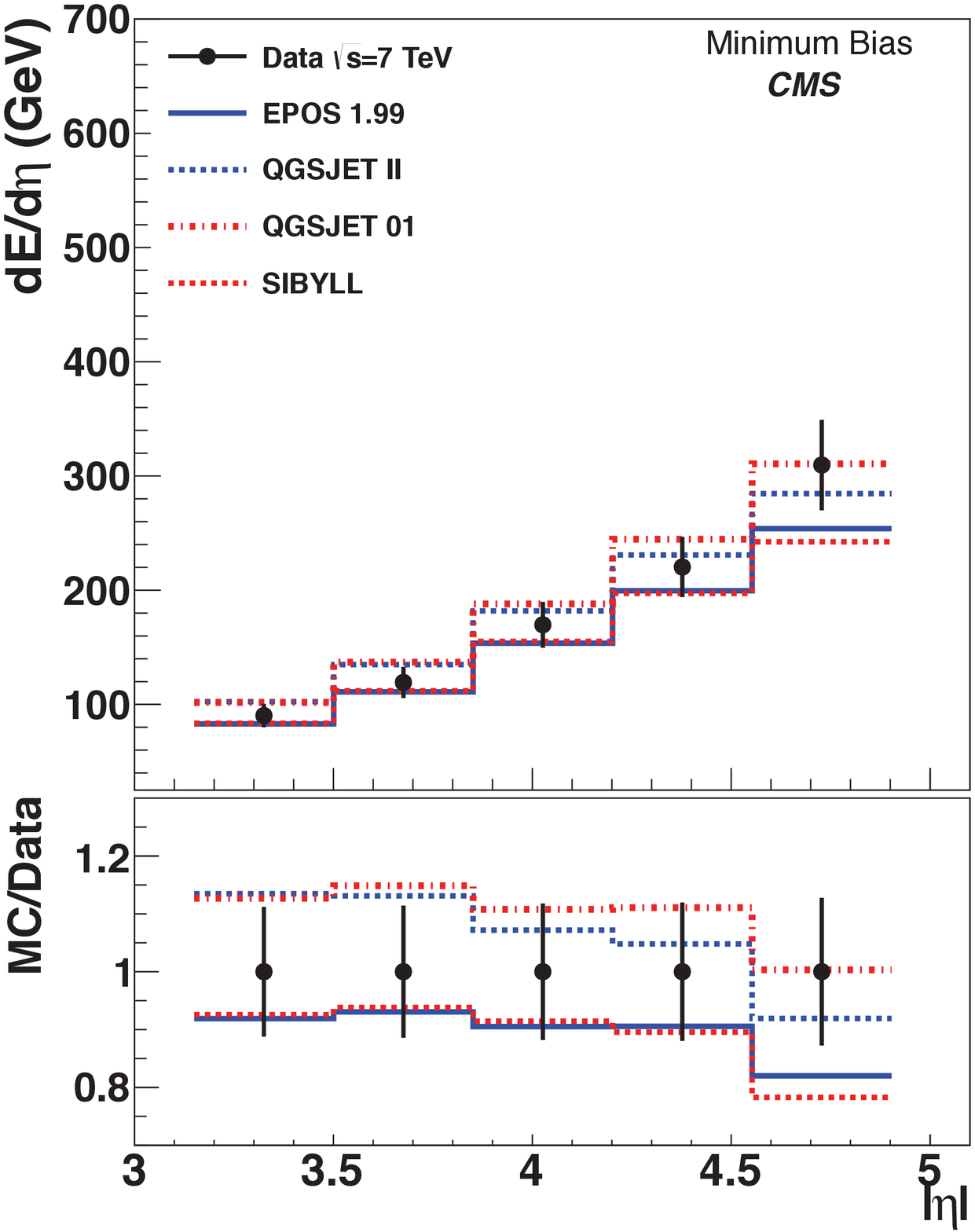}
\end{center}
\caption{Energy flow as function of pseudorapidity in forward direction. CMS data~\cite{Chatrchyan:2011wm} are shown together with model predictions calculated by the CMS Collaboration. The left (right) panel shows the measurements for $900$\,GeV ($7$\,TeV) c.m.s.\ energy.
\label{fig:EnergyFlowCMS}}
\end{figure}

Hence comparisons of model predictions to data from fixed-target and collider experiments are typically made by the authors of the models and using only fully acceptance corrected data. However, over the last years much progress has been made regarding the interaction between the cosmic ray and high energy physics communities. A number of workshops and meetings (including this one) and the direct involvement of cosmic ray physicists in HEP collaborations have lead to a much higher level of awareness and intensified communication of both communities.

Comparisons of high-energy interaction models used for air shower simulations with LHC data can be found in, for example,~\cite{dEnterria:2011kw,Ostapchenko:2012x1,PierogUHECR2012}. Here only some representative examples can be included for illustration. The pseudorapidity distribution of charged particles is shown together with model predictions in Fig.~\ref{fig:EtaLHC}. Not only most central particle distributions are reasonably well bracketed by the model predictions, also the predicted energy flow at larger pseudorapidity is in good agreement with the LHC data. This can be seen in Fig.~\ref{fig:EnergyFlowCMS} where the energy flow measured by CMS~\cite{Chatrchyan:2011wm} is compared to model predictions.
Note that all these model results are true predictions as the models were tuned years before LHC data became available. Nevertheless, in many cases, the minimum bias data are better described by interaction models developed primarily for cosmic ray physics than the various tunes of standard HEP models~\cite{Chatrchyan:2011wm,dEnterria:2011kw}.

\begin{figure}[htb!]
\begin{center}
\includegraphics[width=\textwidth]{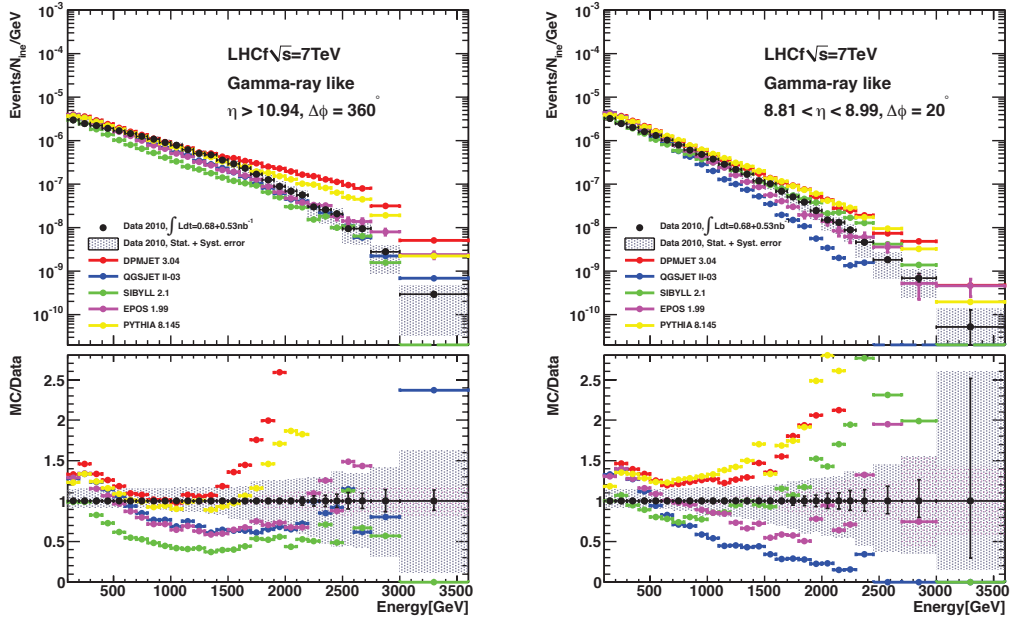}
\end{center}
\caption{Photon spectra in very forward direction as measured by the  LHCf Collaboration~\cite{Adriani:2011nf}. Superimposed are the predictions of hadronic interaction models used for air shower simulations. The lower panels show the relative difference between data and model predictions.
}
\label{fig:LHCf}
\end{figure}
On the other hand, there are also some important distributions measured at LHC that are not well described or bracketed by these pre-LHC interaction models. Examples of large deviations from measurements are multiplicity distributions of charged particles and particle production spectra at large pseudorapidity and Feynman-$x$. The possible impact of these deviations on air shower predictions is subject to ongoing research and not yet fully understood. In Fig.~\ref{fig:LHCf} the Feynman-$x$ distribution of photons produced in forward direction is shown~\cite{Adriani:2011nf}. The data have been obtained within the LHCf experiment that is specifically built for measuring particles in the very forward direction.

In general, the predictions of the different cosmic ray interaction models bracket the LHC data reasonably well~\cite{dEnterria:2011kw}. This observation supports the hope that simulating air showers with different interaction models should also bracket the correct, but unknown predictions for air showers. On the other hand, this success does not mean that there is no need to improve the interaction models. There is not a single interaction model that reproduces most of new LHC data well and, hence, would be much better than the others. Each of the models needs to be re-tuned or underlying ideas be extended to obtain a better description of the data.

Many new data sets measured in LHC experiments and also the fixed-target experiments NA49~\cite{Baatar:2012fua} and NA61~\cite{Abgrall:2011ae,Abgrall:2011ts} became available within the last $2-3$ years and the process of tuning and modifying the interaction models to obtain an improved description is ongoing. At the time of writing these proceedings the new model versions QGSJET~II.04~\cite{Ostapchenko:2012x1} and EPOS LHC~\cite{PierogUHECR2012} have been released as the first versions of interaction models tuned to LHC data. In addition, work is in progress to develop improved versions of SIBYLL and DPMJET that are also tuned to the new LHC data.

Finally it should be mentioned that, in addition to accelerator data, also air shower measurements provide important information on hadronic interactions. For example, cross section measurements made with air shower observables~\cite{UlrichUHECR2012} can be used to estimate not only the rise of the cross section but also to verify the model approximations used to calculate the proton-air cross section from the data on proton-proton cross sections. Even though the systematic uncertainties of air shower measurements of this kind are much higher than that of accelerator data, such measurements are the only way to study particle production at interaction energies well beyond the range of colliders~\cite{Belov:2006mb,Knurenko:1999cr,Knurenko:2011aa,Abreu:2012wt}.


\section{Overall Description of Shower Characteristics\label{sec:ShowerDescription}}

Simulating air showers with a realistic energy spectrum, primary mass composition, and arrival direction distribution, the predicted and observed distributions of the various observables of reconstructed showers can be compared. Such comparisons are very important as they are end-to-end tests of the simulation chain for the experiments. Only if good agreement is found one can cross-check the impact of quality cuts applied in the process of data analysis and avoid unexpected biases. The results of such end-to-end simulations depend, of course, on the assumed mass composition of the primary particles and the employed hadronic interaction models.

Examples of such comparisons for fluorescence detectors can be found in \cite{Abbasi:2004nz,Abbasi:2005ni,Abreu:2010zzl,AbuZayyad:2012qk}. A number of unpublished comparisons of typical observables measured with the Auger and TA surface detector arrays (zenith and azimuth angles, number of stations per event, signal size distribution) have been presented at this meeting. They all show very good agreement between the measured distributions and the corresponding Monte Carlo predictions. While the TA simulations were done for proton primaries, the Auger simulations were based on a $50/50$ proton-iron mixture. Both collaborations used QGSJET II.03 as high-energy interaction model. The TA data are better described by a primary composition of only protons than using $100$\% iron. In case of the Auger simulations, which were also made assuming only proton or iron primaries, it was found that most of the surface detector observables exhibit only a very limited sensitivity to the primary mass composition.


It should be considered as an important success of hadronic interaction models and modern air shower simulation packages that such a good overall description of the general features of the observed showers is reached. However, different combinations of primary mass composition and shower energy can lead to very similar surface detector signals (for example, the signal at $600$ or $1000$\,m in units of that of vertical muons). Therefore it is very important to use additional information to determine, for example, the primary energy in a composition-independent way. This is done with fluorescence telescopes in the case of the Auger Observatory~\cite{Abraham:2009pm} and Telescope Array~\cite{Tokuno:2012mi}, and with non-imaging Cherenkov light measurements in the Yakutsk setup~\cite{Ivanov:2009dx}.


\section{Detailed Comparison with Shower Data\label{sec:Discrepancies}}

Detecting air showers simultaneously with fluorescence telescopes and an array of surface detectors at ground is often referred to as hybrid measurements. One of the first setups of this type was the prototype instrument of the High Resolution Fly's Eye (HiRes)~\cite{AbuZayyad:2000uu} that was operated in coincidence with the CASA-MIA array~\cite{Borione:1994iy} in Utah. Already in this prototype experiment strong indications for a discrepancy between the  measured density of muons at $600$\,m from the shower core and the one expected from simulations were found, if the composition inferred from the measurements of the longitudinal shower profiles was used~\cite{AbuZayyad:1999xa}. While the data on the depth of shower maximum suggested the conclusion that a transition from a heavy to a light, almost proton dominated composition is seen, the density of muons stayed above or at the level of the predictions for iron primaries.

\begin{figure}[htb!]
\begin{center}
\includegraphics[width=0.505\textwidth]{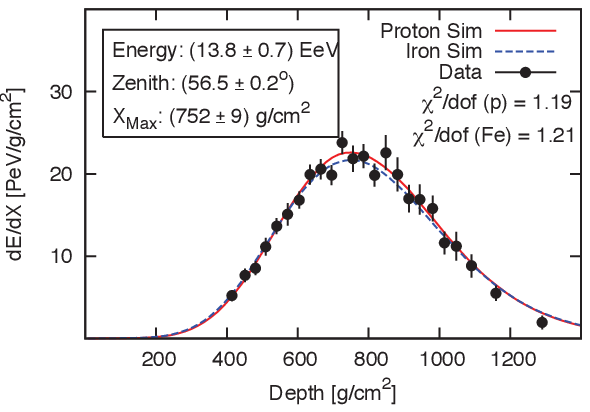}
\includegraphics[width=0.485\textwidth]{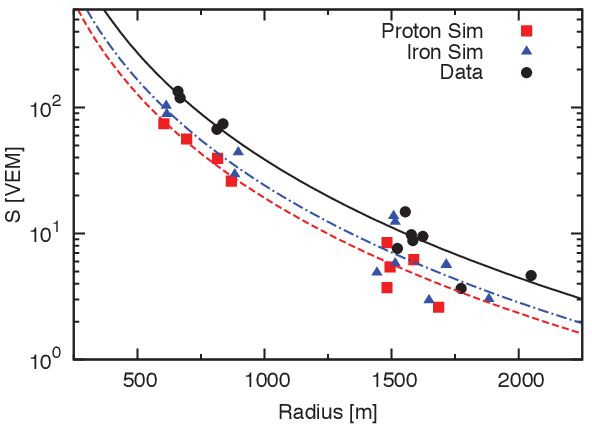}
\end{center}
\caption{
Longitudinal and lateral profiles of one high-energy shower observed simultaneously with the fluorescence and the surface detectors of the Pierre Auger Observatory~\cite{Abreu:2011pe}. The data are compared with simulated showers for proton and iron primaries that predict the same longitudinal profiles. The simulations were made with SENECA and QGSJET II.03/FLUKA as high-/low-energy interaction models.
\label{fig:TopDownAuger}}
\end{figure}

Covering energies higher than those of the early HiRes-MIA measurements, the data of the Auger Collaboration also indicate a discrepancy between the observed and predicted number of muons in air showers~\cite{Abreu:2011pe}. Different methods have been applied to obtain either the expected surface detector signal $S(1000)$ or the muon density at $1000$\,m from the shower core for a given primary energy or even an observed longitudinal shower profile (see \cite{YushkovUHECR2012} for a review given at this meeting). This discrepancy is most directly displayed in a shower-by-shower comparison of simulated profiles with those obtained from hybrid measurements, see Fig.~\ref{fig:TopDownAuger}. Compared to proton showers as reference, the measured values of $S(1000)$ are about 1.5 to 2 times higher that the simulated ones. Studying inclined showers of about $10^{19}$\,eV with zenith angles larger than $60^\circ$, whose particle content at ground is completely dominated by muons, gives a ratio of~\cite{Abreu:2011pe}
\begin{equation}
\left. \frac{N_{\mu, \rm data}}{N_{\mu, \rm MC}}\right|_{\rm QGSJET,p} = 2.13 \pm 0.04 {\rm (stat)} \pm 0.11 {\rm (sys)},
\end{equation}
where the systematic uncertainty due to the $22$\% energy scale uncertainty is not included. This ratio is about $1.2$ for iron showers simulated with EPOS 1.99. The data on inclined showers imply that the observed discrepancy is closely related to, if not dominated by a deficiency in simulating the muon component of air showers.

The detectors of the Auger, TA and Yakutsk air shower arrays are differing in their sensitivities to muons and electromagnetic particles, and the Yakutsk array even contains a number of shielded muon detectors. For example, the contribution of muons to the scintillator signal of the TA surface detectors will never exceed $30$\% for showers with zenith angles less than $45^\circ$. In comparison, a signal fraction of $30 - 80$\% is expected due to muons for the Auger water Cherenkov tanks in the zenith angle range from $0 - 60^\circ$. Therefore similar studies of the TA and Yakutsk Collaborations can provide a very important, independent information on a possible muon deficit of air shower simulations.


The Yakutsk Collaboration compared the predicted and measured charged particle and muon lateral distributions for a number of high energy showers. Within the uncertainties implied by the unknown primary composition, no significant discrepancies between the predictions obtained with both QGSJET II/FLUKA and EPOS/UrQMD and the data are found~\cite{Knurenko:2010et}.


The TA Collaboration studied the difference between the energy one would assign to a shower based on the calorimetric fluorescence light measurement and that derived from the comparison with simulations of the surface detector signal. It was found that the fluorescence-assigned energy is about $27$\% lower than the energy estimated from the comparison with simulated showers~\cite{AbuZayyad:2012ru}. Given the limited sensitivity of the TA surface detector stations, no direct study of the muon contribution to the overall signal has been done so far.


The observations of all three collaborations can be brought into qualitative agreement if the different energy scales of the experiments are taken into account. The factor with which the energy scales need to be re-scaled are listed in the report of the {Spectrum Working Group}~\cite{SpectrumUHECR2012}. For example, for the same shower, the Yakutsk simulations are done at an energy about two times higher than that of the Auger Observatory. Hence, also about a factor two more muons are predicted in the Yakutsk simulations in comparison to Auger simulations. Given that the Yakutsk simulations agree with their measurements, this implies two times more muons in the Yakutsk data than predicted by simulations if the Auger energy scale is used. A similar estimate can be done for the TA-Auger comparison, where one has to keep in mind that only $\sim 20$\% of the scintillator signal stems from muons. If the discrepancy observed by the Auger Collaboration would be attributed to the muon component only, the expected energy rescaling would be somewhat smaller than the $\sim 27$\% reported by the TA Collaboration~\cite{AbuZayyad:2012ru}.

It remains to be shown how much of the apparent discrepancy between the surface detector signals and the predictions based on calorimetric shower energy determination can be cured by adopting different energy scales in the experiments. A reduction of the systematic uncertainties of the energy assignment to air showers will be needed to quantify more reliably possible discrepancies between shower data and predictions.


\section{Hadronic Interactions and Muon Production in Air Showers\label{sec:MuonProduction}}

The observation of a possible muon deficit in simulated showers in comparison to Auger measurements, first reported already in 2007~\cite{Engel:2007cm}, has triggered a number of theoretical studies searching for modifications of hadronic interactions that could result in an enhancement of the muonic shower component. Changes of the inelastic cross section, inelasticity of interactions, secondary particle multiplicity at high energy, and many other parameters lead only to moderate changes of the predicted number of muons (see, for example, \cite{Ulrich:2010rg,Parsons:2011ad}). Also scenarios with new physics processes such as string percolation~\cite{AlvarezMuniz:2012dd} or the drastic change of interaction properties due to, for example, chiral symmetry restoration~\cite{FarrarUHECR2012} have been discussed.

So far all proposed changes that lead to a significant increase of the number of muons in air showers are directly or indirectly based on either one or both of the following effects:

{\bf (i)}  An increase of the production rate of particles that do not decay, for example baryon-antibaryon pairs, leads to higher muon multiplicities of showers since these particles will stay being part of the hadronic shower component and loose their energy only by producing further hadronic particles. This effect has been discussed first in~\cite{Grieder:1973x1} and is one of the differences of the EPOS model with respect to the other models~\cite{Pierog:2006qv}.

{\bf (ii)} A change of the type of the leading particle produced in inelastic interactions can also be very efficient in reducing the energy that is transferred to the electromagnetic shower component~\cite{Drescher:2007hc}. The majority of sub-showers in an extensive air shower are initiated by charged pions. The chance probability of producing a charged pion or a neutral pion as leading particle in charged pion interactions is about $2:1$. Replacing all leading $\pi^0$ by $\rho^0$ mesons, which have the same valence quarks but are spin 1 particles, leads to a drastic enhancement of muon production since neutral $\rho$ mesons decay immediately into two charged pions~\cite{OstapchenkoPrivComm}. Indeed, fixed-target data~\cite{Schouten:1981ts,AguilarBenitez:1989fn,Agababyan:1990df} indicate that, in contrast to conventional model predictions, the production of $\rho^0$ dominates that of $\pi^0$ for Feynman-$x$ larger than $0.5$.

Both effects are the reason for the higher muon multiplicities predicted with EPOS. Leading rho mesons are explicitly generated for pion interactions in the new version II.04 of QGSJET, boosting the muon multiplicity in air showers significantly. Depending on the relative importance of effect (i) and energy distribution of the baryon pairs, the energy spectrum of the additionally produced muons is very soft and might be in contradiction to shower attenuation data.

Finally it should be mentioned that also an increase of the production of kaons could lead to a higher muon multiplicity and also a harder muon energy spectrum~\cite{Drescher:2007hc,AlvarezMuniz:2012dd}.


\section{Conclusions and Outlook}

Over the last two decades the quality and predictive power of air shower simulations has improved significantly and a very good overall description of most of the shower features observed in experiments has been reached.

The predictions of independently developed shower simulation packages agree reasonably well with each other if the same hadronic interaction models are used for the simulations. While there is very good agreement for the electromagnetic shower component, the differences between the predictions  for muon multiplicities and lateral distributions can be as large as $\sim 5 - 10$\%. These differences are most likely related to the use of different low-energy interaction models. A comprehensive comparison of the different shower simulation packages, similar to the recent study of CORSIKA and COSMOS predictions~\cite{Roh:2013ev}, should be made to quantify the systematic uncertainties of the predictions and possibly also to identify and address shortcomings of the simulation packages.

The limited theoretical understanding of modeling hadronic multiparticle production together with the limitations of accelerator measurements in energy, covered phase space, and projectile-target combinations are the dominating source of systematic uncertainties of air shower predictions. Because of this, the systematic uncertainties of the model predictions cannot be estimated reliably. More work is needed to improve QCD calculations in the low-$p_\perp$ domain, in particular to understand screening and saturation effects, and to develop alternative models to describe particle production.

The new LHC data provide extremely useful input for tuning hadronic interaction models. Even though the first LHC data on multiplicities and cross sections were well bracketed by the predictions of interaction models used for air shower simulations, the comparison to data revealed  the need for further model developments and tuning. Improved and re-tuned versions of EPOS and QGSJET are already available and similar versions of DPMJET and SIBYLL are in preparation.

The interaction between the cosmic ray and high energy physics communities has intensified and the direct engagement of cosmic ray physicists in LHC and fixed-target experiments has lead to a much better understanding of the needs of the cosmic ray community. There is also very large interest from the side of LHC communities to have cosmic ray physicists being involved in the analysis of accelerator data. One example of the achieved progress is the use of cosmic ray interaction models by LHC collaborations to compare data with predictions in publications.

Both the Auger and TA Collaborations have found indications for a discrepancy between the expected and observed surface detector signals for showers with fluorescence energy measurement. Accounting for the different energy scales of the Auger, TA, and Yakutsk experiments the observed discrepancies are consistent with each other. Most likely, the dominating sources of the discrepancies are shortcomings of the simulation of muon production in air showers. The systematic uncertainties of the overall energy scales of the experiments of $20 - 30$\% will have to be reduced significantly to be able to combine data from different experiments for a stronger test of the shower predictions.

While the production of electromagnetic particles is clearly dominated by the electromagnetic cascade induced by photons of very high energy due to neutral pion decay early on in the shower evolution, the muonic component receives contributions from all high-energy interactions above $\sim 20$\,GeV lab.\ energy. Different modifications of the simulation of particle production in low- and high-energy interactions have been considered to increase the number of muons in air showers. Progress has been made by understanding that both production of stable or long-lived hadrons such as baryon pairs and modifications to the particle types generated as leading particles are efficient mechanisms to increase the muon multiplicity without changing significantly the longitudinal shower profile. It remains to be shown whether the discrepancies between the observed and predicted surface detector signals for a given shower energy can be resolved by improving low- and high-energy interaction models using conventional physics assumptions or whether these observations indeed indicate a change of the characteristics of hadronic interactions at energies beyond that of LHC.




\begin{thebibliography}{100}

\bibitem{Auger:1939x1}
P.~Auger, P.~Ehrenfest, R.~Maze, Robley, and A.~Fr{\'e}on,
\comment{ Extensive Cosmic-Ray Showers }
Rev. Mod. Phys. 11 (1939) 288--291.

\bibitem{Kieda:2000ky}
D.~B. Kieda, S.~P. Swordy, and S.~P. Wakely,
\comment{ {A high resolution method for measuring cosmic ray composition beyond
  10-TeV} }
Astropart. Phys. 15 (2001) 287--303.

\bibitem{Gaisser:1978kx}
T.~K. Gaisser, R.~J. Protheroe, K.~E. Turver, and T.~J.~L. McComb,
\comment{ COSMIC RAY SHOWERS AND PARTICLE PHYSICS AT ENERGIES 10**15- EV -  10**18-EV }
Rev. Mod. Phys. 50 (1978) 859--880.

\bibitem{Knapp:2002vs}
J.~Knapp, D.~Heck, S.~J. Sciutto, M.~T. Dova, and M.~Risse,
\comment{ Extensive air shower simulations at the highest energies }
Astropart. Phys. 19 (2003) 77--99.

\bibitem{Anchordoqui:2004xb}
L.~Anchordoqui {\it et~al.},
\comment{ High energy physics in the atmosphere: Phenomenology of cosmic ray
  air showers }
Ann. Phys. 314 (2004) 145--207
 and hep-ph/0407020.

\bibitem{Engel:2011zz}
R.~Engel, D.~Heck, and T.~Pierog,
\comment{ {Extensive air showers and hadronic interactions at high energy} }
Ann. Rev. Nucl. Part. Sci. 61 (2011) 467--489.

\bibitem{Matthews:2005sd}
J.~Matthews,
\comment{ A Heitler model of extensive air showers }
Astropart. Phys. 22 (2005) 387--397.

\bibitem{Ulrich-MPI:2011x1}
R.~Ulrich,
talk given at the Workshop on Multi Parton Interactions (MPI) 2011, Hamburg,
2012.

\bibitem{Bergmann:2006yz}
T.~Bergmann {\it et~al.},
\comment{ One-dimensional hybrid approach to extensive air shower simulation }
Astropart. Phys. 26 (2007) 420--432
 and astro-ph/0606564.

\bibitem{Drescher:2002vp}
H.-J. Drescher and G.~R. Farrar,
\comment{ Dominant contributions to lateral distribution functions in
  ultra-high energy cosmic ray air showers }
Astropart. Phys. 19 (2003) 235--244
 and hep-ph/0206112.

\bibitem{Meurer:2005dt}
C.~Meurer, J.~Bl{\"u}mer, R.~Engel, A.~Haungs, and M.~Roth,
\comment{ Muon production in extensive air showers and its relation to hadronic
  interactions }
Czech. J. Phys. 56 (2006) A211
 and astro-ph/0512536.

\bibitem{Hillas:1982vn}
A.~M. Hillas,
\comment{ {Angular and energy distributions of charged particles in electron
  photon cascades in air} }
J. Phys. G8 (1982) 1461--1473.

\bibitem{Giller:2004cf}
M.~Giller, G.~Wieczorek, A.~Kacperczyk, H.~Stojek, and W.~Tkaczyk,
\comment{ {Energy spectra of electrons in the extensive air showers of
  ultra-high energy} }
J. Phys. G30 (2004) 97--105.

\bibitem{Giller:2005gp}
M.~Giller, H.~Stojek, and G.~Wieczorek,
\comment{ {Extensive air shower characteristics as functions of shower age} }
Int. J. Mod. Phys. A20 (2005) 6821--6824.

\bibitem{Nerling:2006yt}
F.~Nerling, J.~{Bl\"umer}, R.~Engel, and M.~Risse,
\comment{ Universality of electron distributions in high-energy air showers:
  Description of Cherenkov light production }
Astropart. Phys. 24 (2006) 421--437
 and astro-ph/0506729.

\bibitem{Lafebre:2009en}
S.~Lafebre, R.~Engel, H.~Falcke, J.~H{\"o}randel, T.~Huege, J.~Kuijpers, and
  R.~Ulrich,
\comment{ {Universality of electron-positron distributions in extensive air
  showers} }
Astropart. Phys. 31 (2009) 243--254
 and arXiv:0902.0548 [astro-ph.HE].

\bibitem{Lipari:2008td}
P.~Lipari,
\comment{ {The Concepts of 'Age' and 'Universality' in Cosmic Ray Showers} }
Phys. Rev. 79 (2008) 063001
 and arXiv:0809.0190 [astro-ph].

\bibitem{Baltrusaitis:1985mx}
R.~M. Baltrusaitis {\it et~al.}  (Fly's Eye Collab.),
\comment{ The Utah Fly's Eye Detector }
Nucl. Instrum. Meth. A240 (1985) 410--428.

\bibitem{Andringa:2011ik}
S.~Andringa, L.~Cazon, R.~{Concei\c{c}\~ao}, and M.~Pimenta,
\comment{ {The Muonic longitudinal shower profiles at production} }
Astropart. Phys. 35 (2012) 821--827
 and arXiv:1111.1424 [hep-ph].

\bibitem{Cazon:2012ti}
L.~Cazon, R.~{Concei\c{c}\~ao}, M.~Pimenta, and E.~Santos,
\comment{ {A model for the transport of muons in extensive air showers} }
Astropart. Phys. 36 (2012) 211--223
 and arXiv:1201.5294 [astro-ph.HE].

\bibitem{Schmidt:2007vq}
F.~Schmidt, M.~Ave, L.~Cazon, and A.~S. Chou,
\comment{ {A Model-Independent Method of Determining Energy Scale and Muon
  Number in Cosmic Ray Surface Detectors} }
Astropart. Phys. 29 (2008) 355--365
 and arXiv:0712.3750 [astro-ph].

\bibitem{Schmidt:2009ge}
F.~Schmidt {\it et~al.}  (Pierre Auger Collab.),
\comment{ {Measurements of the Muon Content of UHECR Air Showers with the
  Pierre Auger Observatory} }
Nucl. Phys. Proc. Suppl. 196 (2009) 141--146
 and arXiv:0902.4613 [astro-ph.HE].

\bibitem{Ave:2011x1}
M.~Ave, R.~Engel, J.~Gonzalez, D.~Heck, T.~Pierog, and M.~Roth,
\comment{ Extensive Air Shower Universality of Ground Particle Distributions }
Proc. of 31th Int. Cosmic Ray Conf., Beijing  (2011) \#1025.

\bibitem{Maris:2009x1}
I.~C. Maris {\it et~al.}  (NA61 Collab.),
\comment{ Hadron Production Measurements with the NA61-SHINE Experiment and
  their Relevance for Air Shower Simulations }
Proc of 31th Int. Cosmic Ray Conf., {\L}\'{o}d\'{z}  (2009) .

\bibitem{Sciutto:AIRES}
S.~J. Sciutto,
\comment{ AIRES: A system for air shower simulations (version 2.2.0) }
astro-ph/9911331 and
astro-ph/0106044.

\bibitem{Heck98a}
D.~Heck, J.~Knapp, J.~Capdevielle, G.~Schatz, and T.~Thouw,
\comment{ CORSIKA: a Monte Carlo code to simulate extensive air showers }
Wissenschaftliche Berichte, Forschungszentrum Karlsruhe FZKA 6019 (1998) .

\bibitem{cosmos}
K.~Kasahara {\it et~al.},
\comment{ COSMOS } http://cosmos.n.kanagawa-u.ac.jp/cosmosHome.

\bibitem{Hillas81a}
A.~M. Hillas,
\comment{ Two Interesting Techniques for Monte Carlo Simulation of Very High
  Energy Hadron Cascades }
Proc. of 17th Int. Cosmic Ray Conf., Paris 8 (1981) 193.

\bibitem{Kobal:2001jx}
M.~Kobal,
\comment{ A thinning method using weight limitation for air-shower simulations
  }
Astropart. Phys. 15 (2001) 259--273.

\bibitem{Gorbunov:2007vj}
D.~S. Gorbunov, G.~I. Rubtsov, and S.~V. Troitsky,
\comment{ Air-shower simulations with and without thinning: Artificial
  fluctuations and their suppression }
Phys. Rev. D76 (2007) 043004
 and astro-ph/0703546.

\bibitem{Hansen:2010uk}
P.~Hansen, J.~Alvarez-Muniz, and R.~Vazquez,
\comment{ {A comprehensive study of shower to shower fluctuations} }
Astropart. Phys. 34 (2011) 503--512
 and arXiv:1004.3666 [astro-ph.HE].

\bibitem{Billoir:2008zz}
P.~Billoir,
\comment{ {A sampling procedure to regenerate particles in a ground detector
  from a 'thinned' air shower simulation output} }
Astropart. Phys. 30 (2008) 270--285.

\bibitem{Stokes:2011wf}
B.~Stokes, R.~Cady, D.~Ivanov, J.~Matthews, and G.~Thomson,
\comment{ {Dethinning Extensive Air Shower Simulations} }
Astropart. Phys. 35 (2012) 759--766
 and arXiv:1104.3182 [astro-ph.IM].

\bibitem{Stokes:2011xk}
B.~Stokes, R.~Cady, D.~Ivanov, J.~Matthews, and G.~Thomson,
\comment{ {A Simple Parallelization Scheme for Extensive Air Shower
  Simulations} }
arXiv:1103.4643 [astro-ph.IM].

\bibitem{Drescher:2002cr}
H.-J. Drescher and G.~R. Farrar,
\comment{ Air shower simulations in a hybrid approach using cascade equations }
Phys. Rev. D67 (2003) 116001
 and astro-ph/0212018.

\bibitem{Hillas:1997tf}
A.~M. Hillas,
\comment{ {Shower simulation: Lessons from MOCCA} }
Nucl. Phys. Proc. Suppl. 52B (1997) 29--42.

\bibitem{egs4}
W.~Nelson {\it et~al.},
\comment{ The EGS4 Code System }
SLAC-265, Stanford Linear Accelerator Center  (1985) .

\bibitem{Cillis:1998hf}
A.~N. Cillis, H.~Fanchiotti, C.~A. Garcia~Canal, and S.~J. Sciutto,
\comment{ Influence of the LPM effect and dielectric suppression on particle
  air showers }
Phys. Rev. D59 (1999) 113012
 and astro-ph/9809334.

\bibitem{Fesefeldt85a}
H.~Fesefeldt,
\comment{ Simulation of hadronic showers - Physics and applications } preprint
  PITHA-85/02, RWTH Aachen,
1985.

\bibitem{Ferrari:2005zk}
A.~Ferrari, P.~R. Sala, A.~Fasso, and J.~Ranft,
\comment{ {FLUKA: A multi-particle transport code (Program version 2005)} }
CERN-2005-010.

\bibitem{Bleicher99a}
M.~Bleicher {\it et~al.},
\comment{ Relativistic hadron-hadron collisions in the ultra-relativistic
  quantum molecular dynamics model }
J. Phys. G: Nucl. Part. Phys. 25 (1999) 1859.

\bibitem{Ranft:1994fd}
J.~Ranft,
\comment{ The Dual parton model at cosmic ray energies }
Phys. Rev. D51 (1995) 64--84.

\bibitem{Roesler02a}
S.~Roesler, R.~Engel, and J.~Ranft,
\comment{ The Monte Carlo event generator DPMJET-III at cosmic ray energies }
Proc of 27th Int. Cosmic Ray Conf., Hamburg 2 (2002) 439.

\bibitem{Bopp:2005cr}
F.~W. Bopp, J.~Ranft, R.~Engel, and S.~Roesler,
\comment{ Antiparticle to Particle Production Ratios in Hadron-Hadron and d-Au
  Collisions in the DPMJET-III Monte Carlo }
Phys. Rev. C77 (2008) 014904
 and hep-ph/0505035.

\bibitem{Werner:2005jf}
K.~Werner, F.-M. Liu, and T.~Pierog,
\comment{ Parton ladder splitting and the rapidity dependence of transverse
  momentum spectra in deuteron gold collisions at RHIC }
Phys. Rev. C74 (2006) 044902
 and hep-ph/0506232.

\bibitem{Pierog:2009zt}
T.~Pierog and K.~Werner,
\comment{ {EPOS Model and Ultra High Energy Cosmic Rays} }
Nucl. Phys. Proc. Suppl. 196 (2009) 102--105
 and arXiv:0905.1198 [hep-ph].

\bibitem{Werner:2010ss}
K.~Werner, I.~Karpenko, and T.~Pierog,
\comment{ {The 'Ridge' in Proton-Proton Scattering at 7 TeV} }
Phys. Rev. Lett. 106 (2011) 122004
 and arXiv:1011.0375 [hep-ph].

\bibitem{Kalmykov:1997te}
N.~N. Kalmykov, S.~S. Ostapchenko, and A.~I. Pavlov,
\comment{ Quark-gluon string model and EAS simulation problems at ultra-high
  energies }
Nucl. Phys. Proc. Suppl. 52B (1997) 17--28.

\bibitem{Kalmykov:1993qe}
N.~N. Kalmykov and S.~S. Ostapchenko,
\comment{ The Nucleus-nucleus interaction, nuclear fragmentation, and
  fluctuations of extensive air showers }
Phys. Atom. Nucl. 56 (1993) 346--353.

\bibitem{Ostapchenko:2005nj}
S.~Ostapchenko,
\comment{ Non-linear screening effects in high energy hadronic interactions }
Phys. Rev. D74 (2006) 014026
 and hep-ph/0505259.

\bibitem{Ostapchenko:2004ss}
S.~Ostapchenko,
\comment{ QGSJET-II: Towards reliable description of very high energy hadronic
  interactions }
Nucl. Phys. Proc. Suppl. 151 (2006) 143--146
 and hep-ph/0412332.

\bibitem{Ostapchenko:2007qb}
S.~Ostapchenko,
\comment{ Status of QGSJET }
AIP Conf. Proc. 928 (2007) 118--125
 and arXiv:0706.3784 [hep-ph].

\bibitem{Ahn:2009wx}
E.-J. Ahn, R.~Engel, T.~K. Gaisser, P.~Lipari, and T.~Stanev,
\comment{ {Cosmic ray interaction event generator SIBYLL 2.1} }
Phys. Rev. D 80 (2009) 094003
 and arXiv:0906.4113 [hep-ph].

\bibitem{Engel:1992vf}
J.~Engel, T.~K. Gaisser, T.~Stanev, and P.~Lipari,
\comment{ Nucleus-nucleus collisions and interpretation of cosmic ray cascades
  }
Phys. Rev. D46 (1992) 5013--5025.

\bibitem{Fletcher:1994bd}
R.~S. Fletcher, T.~K. Gaisser, P.~Lipari, and T.~Stanev,
\comment{ SIBYLL: An Event generator for simulation of high-energy cosmic ray
  cascades }
Phys. Rev. D50 (1994) 5710--5731.

\bibitem{Heck:2002yf}
D.~Heck, M.~Risse, and J.~Knapp,
\comment{ Comparison of hadronic interaction models at Auger energies }
Nucl. Phys. Proc. Suppl. 122 (2003) 364--367
 and astro-ph/0210392.

\bibitem{Heck:2004rq}
D.~Heck,
\comment{ {Low energy hadronic interaction models} }
Nucl. Phys. Proc. Suppl. 151 (2006) 127--134
 and astro-ph/0410735.

\bibitem{Heck08CORSIKASchool}
D.~Heck,
\comment{ The Influence of Hadronic Interaction Models on Simulated
  Air-Showers: A Phenomenologic Comparison } Talk given at CORSIKA School 2008,
  Freudenstadt, Germany, 25 - 30 November 2008, available at
  http://www-ik.fzk.de/corsika/corsika-school2008/,
2008.

\bibitem{Kim:2011x1}
J.~Kim {\it et~al.}  (TA Collab.),
\comment{ Comparison Study of Extensive Air Shower Simulations with COSMOS and
  CORSIKA }
Proc of 32th Int. Cosmic Ray Conf., Beijing  (2011) 0812.

\bibitem{Roh:2013ev}
S.~Roh {\it et~al.},
\comment{ {A comparison study of CORSIKA and COSMOS simulations for extensive
  air showers} }
Astropart. Phys. 44 (2013) 1--8
 and arXiv:1301.5060 [astro-ph.HE].

\bibitem{Ortiz:2004gb}
J.~A. Ortiz, G.~A. Medina~Tanco, and V.~de~Souza,
\comment{ {Longitudinal development of extensive air showers: Hybrid code
  SENECA and full Monte Carlo} }
Astropart. Phys. 23 (2005) 463--476
 and astro-ph/0411421.

\bibitem{Allen:2009x1}
J.~Allen and G.~R. Farrar,
\comment{ Detailed comparison of the SENECA and CORSIKA shower simulation
  packages }
Proc of 31th Int. Cosmic Ray Conf., {\L}\'{o}d\'{z}  (2009) 1557.

\bibitem{ToderoPeixoto:2013x1}
C.~J. {Todero Peixoto}, V.~{de Souza}, and J.~Bellido,
\comment{ {Comparison of the moments of the Xmax distribution predicted by
  different cosmic ray shower simulation models} }
arXiv:1301.5555 [astro-ph.HE].

\bibitem{Khachatryan:2010us}
V.~Khachatryan {\it et~al.}  (CMS Collab.),
\comment{ {Transverse-momentum and pseudorapidity distributions of charged
  hadrons in pp collisions at sqrt(s) = 7 TeV} }
Phys. Rev. Lett. 105 (2010) 022002
 and arXiv:1005.3299 [hep-ex].

\bibitem{Aad:2010ac}
G.~Aad {\it et~al.}  (ATLAS Collab.),
\comment{ {Charged-particle multiplicities in pp interactions measured with the
  ATLAS detector at the LHC} }
New J. Phys. 13 (2011) 053033
 and arXiv:1012.5104 [hep-ex].

\bibitem{Aamodt:2010pp}
K.~Aamodt {\it et~al.}  (ALICE Collab.),
\comment{ {Charged-particle multiplicity measurement in proton-proton
  collisions at sqrt(s) = 7 TeV with ALICE at LHC} }
Eur. Phys. J. C68 (2010) 345--354
 and arXiv:1004.3514 [hep-ex].

\bibitem{Aamodt:2010ft}
K.~Aamodt {\it et~al.}  (ALICE Collab.),
\comment{ {Charged-particle multiplicity measurement in proton-proton
  collisions at sqrt(s) = 0.9 and 2.36 TeV with ALICE at LHC} }
Eur. Phys. J. C68 (2010) 89--108
 and arXiv:1004.3034 [hep-ex].

\bibitem{Chatrchyan:2011wm}
S.~Chatrchyan {\it et~al.}  (CMS Collab.),
\comment{ {Measurement of energy flow at large pseudorapidities in $pp$
  collisions at $\sqrt{s} = 0.9$ and 7 TeV} }
JHEP 1111 (2011) 148
 and arXiv:1110.0211 [hep-ex].

\bibitem{dEnterria:2011kw}
D.~{d'Enterria}, R.~Engel, T.~Pierog, S.~Ostapchenko, and K.~Werner,
\comment{ {Constraints from the first LHC data on hadronic event generators for
  ultra-high energy cosmic-ray physics} }
Astropart. Phys. 35 (2011) 98--113
 and arXiv:1101.5596 [astro-ph.HE].

\bibitem{Ostapchenko:2012x1}
S.~Ostapchenko,
\comment{ Studies of very high energy cosmic rays: Status, puzzles, and the
  impact of LHC data }
Prog. Theor. Phys. Suppl. 193 (2012) 204--211.

\bibitem{PierogUHECR2012}
T.~Pierog,
these proceedings and talk given at Int. Symposium on Very High Energy Cosmic
  Ray Interactions (ISVHECRI 2012) Berlin,
2012.

\bibitem{Adriani:2011nf}
O.~Adriani {\it et~al.}  (LHCf Collab.),
\comment{ {Measurement of zero degree single photon energy spectra for
  $\sqrt{s}$ = 7\,TeV proton-proton collisions at LHC} }
Phys. Lett. B703 (2011) 128--134
 and arXiv:1104.5294 [hep-ex].

\bibitem{Baatar:2012fua}
B.~Baatar {\it et~al.}  (NA49 Collab.),
\comment{ {Inclusive production of protons, anti-protons, neutrons, deuterons
  and tritons in p+C collisions at 158 GeV/c beam momentum} }
arXiv:1207.6520 [hep-ex].

\bibitem{Abgrall:2011ae}
N.~Abgrall {\it et~al.}  (NA61/SHINE Collab.),
\comment{ {Measurements of Cross Sections and Charged Pion Spectra in
  Proton-Carbon Interactions at 31 GeV/c} }
Phys. Rev. C 84 (2011) 034604
 and arXiv:1102.0983 [hep-ex].

\bibitem{Abgrall:2011ts}
N.~Abgrall {\it et~al.}  (NA61/SHINE Collab.),
\comment{ {Measurement of Production Properties of Positively Charged Kaons in
  Proton-Carbon Interactions at 31 GeV/c} }
Phys. Rev. C85 (2012) 035210
 and arXiv:1112.0150 [hep-ex].

\bibitem{UlrichUHECR2012}
R.~Ulrich {\it et~al.}  (Pierre Auger Collab.),
these proceedings,
2012.

\bibitem{Belov:2006mb}
K.~Belov  (HiRes Collab.),
\comment{ p-air cross-section measurement at 10**18.5-eV }
Nucl. Phys. Proc. Suppl. 151 (2006) 197--204.

\bibitem{Knurenko:1999cr}
S.~P. Knurenko, V.~R. Sleptsova, I.~E. Sleptsov, N.~N. Kalmykov, and S.~S.
  Ostapchenko,
\comment{ Longitudinal EAS development at E(0) = 10**18-eV to 3 x 10**19-eV and
  the QGSJET model }
Proc. of 26th Int. Cosmic Ray Conf., Salt Lake City 1 (1999) 372.

\bibitem{Knurenko:2011aa}
S.~Knurenko and A.~Sabourov,
\comment{ {Fluctuations of the depth of maximum in extensive air showers and
  cross-section of p-air inelastic interaction for energy range 10^15-10^17 eV}
  }
arXiv:1112.2448 [astro-ph.HE].

\bibitem{Abreu:2012wt}
P.~Abreu {\it et~al.}  (Pierre Auger Collab.),
\comment{ {Measurement of the proton-air cross-section at $\sqrt{s}=57$ TeV
  with the Pierre Auger Observatory} }
Phys. Rev. Lett. 109 (2012) 062002
 and arXiv:1208.1520 [hep-ex].

\bibitem{Abbasi:2004nz}
R.~U. Abbasi {\it et~al.}  (HiRes Collab.),
\comment{ A study of the composition of ultra high energy cosmic rays using the
  High Resolution Fly's Eye }
Astrophys. J. 622 (2005) 910--926
 and astro-ph/0407622.

\bibitem{Abbasi:2005ni}
R.~U. Abbasi {\it et~al.}  (HiRes Collab.),
\comment{ Observation of the ankle and evidence for a high-energy break in the
  cosmic ray spectrum }
Phys. Lett. B619 (2005) 271--280
 and astro-ph/0501317.

\bibitem{Abreu:2010zzl}
P.~Abreu {\it et~al.}  (Pierre Auger Collab.),
\comment{ {The exposure of the hybrid detector of the Pierre Auger Observatory}
  }
Astropart. Phys. 34 (2011) 368--381
 and arXiv:1010.6162 [astro-ph.HE].

\bibitem{AbuZayyad:2012qk}
T.~Abu-Zayyad {\it et~al.}  (TA
  Collab.),
\comment{ {The Energy Spectrum of Telescope Array's Middle Drum Detector and
  the Direct Comparison to the High Resolution Fly's Eye Experiment} }
Astropart. Phys. 39-40 (2012) 109--119
 and arXiv:1202.5141 [astro-ph.IM].

\bibitem{Abraham:2009pm}
J.~A. Abraham {\it et~al.}  (Pierre Auger Collab.),
\comment{ {The Fluorescence Detector of the Pierre Auger Observatory} }
Nucl. Instrum. Meth. A620 (2010) 227--251
 and arXiv:0907.4282.

\bibitem{Tokuno:2012mi}
H.~Tokuno {\it et~al.}  (TA
  Collab.),
\comment{ {New air fluorescence detectors employed in the Telescope Array
  experiment} }
Nucl. Instrum. Meth. A676 (2012) 54--65
 and arXiv:1201.0002 [astro-ph.IM].

\bibitem{Ivanov:2009dx}
A.~A. Ivanov, S.~P. Knurenko, and I.~Y. Sleptsov,
\comment{ {Measuring extensive air showers with Cherenkov light detectors of
  the Yakutsk array: The energy spectrum of cosmic rays} }
New J. Phys. 11 (2009) 065008
 and arXiv:0902.1016 [astro-ph.HE].

\bibitem{AbuZayyad:2000uu}
T.~Abu-Zayyad {\it et~al.}  (HiRes Collab.),
\comment{ The prototype high-resolution Fly's Eye cosmic ray detector }
Nucl. Instrum. Meth. A450 (2000) 253--269.

\bibitem{Borione:1994iy}
A.~Borione {\it et~al.}  (CASA-MIA Collab.),
\comment{ A Large air shower array to search for astrophysical sources emitting
  gamma-rays with energies >= 10**14-eV }
Nucl. Instrum. Meth. A346 (1994) 329--352.

\bibitem{AbuZayyad:1999xa}
T.~Abu-Zayyad {\it et~al.}  (HiRes-MIA Collab.),
\comment{ Evidence for changing of cosmic ray composition between $10^{17}$ and
  $10^{18}$ eV from multicomponent measurements }
Phys. Rev. Lett. 84 (2000) 4276
 and astro-ph/9911144.

\bibitem{Abreu:2011pe}
P.~Abreu {\it et~al.}  (Pierre Auger Collab.),
\comment{ {The Pierre Auger Observatory II: Studies of Cosmic Ray Composition
  and Hadronic Interaction models} }
Proc of 32th Int. Cosmic Ray Conf., Beijing  (2011)
 and arXiv:1107.4804.

\bibitem{YushkovUHECR2012}
A.~Yushkov {\it et~al.}  (Pierre Auger Collab.),
these proceedings,
2012.

\bibitem{Knurenko:2010et}
S.~Knurenko, A.~Makarov, M.~Pravdin, and A.~Sabourov,
\comment{ {The Relation Between Charged Particles and Muons With Threshold
  Energy 1 GeV in Extensive Air Showers Registered at the Yakutsk EAS Array} }
arXiv:1010.1182 [astro-ph.HE].

\bibitem{AbuZayyad:2012ru}
T.~Abu-Zayyad, {\it et~al.} (TA Collab.),
\comment{ {The Cosmic Ray Energy Spectrum Observed with the Surface Detector of
  the Telescope Array Experiment} }
Astrophys. J. 768 (2013) L1
 and arXiv:1205.5067 [astro-ph.HE].

\bibitem{SpectrumUHECR2012}
Y.~Tsunesada {\it et~al.},
Report of the Spectrum Working Group, these proceedings,
2012.

\bibitem{Engel:2007cm}
R.~Engel  (Pierre Auger Collab.),
\comment{ Test of hadronic interaction models with data from the Pierre Auger
  Observatory }
Proc. of 30th Int. Cosmic Ray Conf., Merida 4 (2007) 385
 and arXiv:0706.1921 [astro-ph].

\bibitem{Ulrich:2010rg}
R.~Ulrich, R.~Engel, and M.~Unger,
\comment{ {Hadronic Multiparticle Production at Ultra-High Energies and
  Extensive Air Showers} }
Phys. Rev. D83 (2011) 054026
 and arXiv:1010.4310 [hep-ph].

\bibitem{Parsons:2011ad}
R.~Parsons, C.~Bleve, S.~Ostapchenko, and J.~Knapp,
\comment{ {Systematic uncertainties in air shower measurements from high-energy
  hadronic interaction models} }
Astropart. Phys. 34 (2011) 832--839
 and arXiv:1102.4603 [astro-ph.HE].

\bibitem{AlvarezMuniz:2012dd}
J.~Alvarez-Muniz
  {\it et~al.},
\comment{ {Muon production and string percolation effects in cosmic rays at the
  highest energies} }
arXiv:1209.6474 [hep-ph].

\bibitem{FarrarUHECR2012}
G.~Farrar and J.~Allen,
talk given at this symposium,
2012.

\bibitem{Grieder:1973x1}
P.~K.~F. Grieder,
\comment{ The effect of $n\bar n$-production on particle spectra in vertically
  incident and inclined showers derived from simulations }
Proc. of 13th Int. Cosmic Ray Conf., Denver 4 (1973) 2467.

\bibitem{Pierog:2006qv}
T.~Pierog and K.~Werner,
\comment{ Muon Production in Extended Air Shower Simulations }
Phys. Rev. Lett. 101 (2008) 171101
 and astro-ph/0611311.

\bibitem{Drescher:2007hc}
H.-J. Drescher,
\comment{ Remnant Break-up and Muon Production in Cosmic Ray Air Showers }
Phys. Rev. D77 (2007) 056003
 and arXiv:0712.1517 [hep-ph].

\bibitem{OstapchenkoPrivComm}
S.~Ostapchenko,
talk given at Int. Symposium on Very High Energy Cosmic Ray Interactions
  (ISVHECRI 2012) Berlin and private communication,
2012.

\bibitem{Schouten:1981ts}
M.~Schouten {\it et~al.},
\comment{ {INCLUSIVE AND SEMIINCLUSIVE RHO0 PRODUCTION IN PI+ / PI- / K+ / P P
  INTERACTIONS AT 147-GeV/c} }
Z. Phys. C9 (1981) 93--104.

\bibitem{AguilarBenitez:1989fn}
M.~Aguilar-Benitez {\it et~al.}  (LEBC-EHS Collab.),
\comment{ {VECTOR MESON PRODUCTION IN pi- p INTERACTIONS AT 360-GeV/c} }
Z. Phys. C44 (1989) 531.

\bibitem{Agababyan:1990df}
N.~Agababyan {\it et~al.}  (EHS/NA22 Collab.),
\comment{ {Inclusive production of vector mesons in pi+ p interactions at
  250-GeV/c} }
Z. Phys. C46 (1990) 387--395.

\end{thebibliography}

\end{document}